# Conservation of angular momentum in second harmonic generation from under-dense plasmas

Chen-Kang Huang*✉, Chaojie Zhang, Zan Nie, Kenneth A. Marsh, Chris E. Clayton & Chandrashekhar Joshi*✉

University of California Los Angeles, Los Angeles, CA 90095, USA

✉Email: ckhuang@ucla.edu; cjoshi@ucla.edu

## Abstract

Spin and orbital angular momentum of an optical beam are two independent parameters that exhibit distinct effects on mechanical objects. However, when laser beams with angular momentum interact with plasmas, one can observe the interplay between the spin and the orbital angular momentum. Here, by measuring the helical phase of the second harmonic $2\omega$ radiation generated in an underdense plasma using a known spin and orbital angular momentum pump beam, we verify that the total angular momentum of photons is conserved and observe the conversion of spin to orbital angular momentum. We further determine the source of the $2\omega$ photons by analyzing near field intensity distributions of the $2\omega$ light. The $2\omega$ images are consistent with these photons being generated near the largest intensity gradients of the pump beam in the plasma as predicted by the combined effect of spin and orbital angular momentum when Laguerre-Gaussian beams are used.

A paraxial light beam can carry two types of angular momentum, commonly called spin angular momentum (SAM) and orbital angular momentum (OAM). The former is due to the beam's polarization state while the latter arises from its spatial mode [1]. Such beams are known to produce distinct effects on mechanical objects. For instance, a circularly polarized beam is known to exert a torque on a suspended birefringent disc [2]. More recently, vortex beams with helical phase front have been shown to induce rotation of trapped particles [3, 4]. These effects are the result of angular momentum transfer from light to microscopic objects. While the beam polarization is associated with the rotation of particles about their own axis, the helical phase front makes particles rotate about the beam axis. Independent but simultaneous effects of spin and orbital angular momentum have been used in sophisticated manipulation of nanoparticles [5, 6, 7] and enable numerous applications [8, 9]. When an inhomogeneous medium retains no angular momentum transferred from the beam, the coupling of the two types of angular momentum could lead to the conversion between SAM and OAM. Such effect has been observed in space-variant gratings [10], liquid crystals [11], and plasmonic metasurfaces [12, 13].

Unlike the interaction between SAM/OAM light and neutral matter, the interaction between such light and plasma is less studied, although the latter too is expected to be governed by the conservation laws for energy, momentum, and angular momentum. While the transfer of energy and linear momentum in laser-plasma interactions has been extensively studied [14, 15], there are but a handful of theoretical studies on how the angular momentum is transferred from a laser beam to either the plasma electrons or secondary photons that may be produced during laser-plasma interaction. For instance, it has been theoretically shown that OAM light can be effectively coupled to electron plasma/ion acoustic waves through stimulated Raman and Brillouin (parametric) instabilities [16]. It has also been theoretically demonstrated that the angular momentum of light can be transferred to a plasma wakefield using a spatiotemporally shaped beam [17]. There are proposals to use intense OAM laser pulses for creating a suitable wake for electron/positron acceleration [18, 19, 20], ion acceleration [21], radiation emission [22], and magnetic field generation [23, 24]. Most of these applications require vortex beams with relativistic intensity.

Using a plasma medium to generate vortex beams is tempting since plasmas have no damage threshold, although in this work efficient vortex beam generation is not principal concern. Various ideas have been proposed to generate vortex beams using plasmas, such as spin-to-orbital angular momentum conversion [25, 26], high harmonic generation from laser interaction using solid target plasmas [27, 28], and stimulated Raman scattering [29]. Despite these proposals, there are few experiments [21, 28, 30] due to the highly nonlinear nature of these processes and the experimental difficulties associated with aberrations of large-aperture vortex beams [31]. Through an experimenter's point of view, the process of second harmonic generation (SHG) in plasma as a platform for spin-to-orbital angular momentum conversion and testing the conservation law for the optical angular momentum is attractive since it is arguably a well understood nonlinear process. By using finite diameter plasmas with sharp density gradients or by using tightly focused intense pulses that carry SAM and/or OAM it is possible to generate second harmonic vortex beams. It is not clear which of these two mechanisms is dominant in a given experiment. For example, Gordan et al. [25] proposed a scheme to generate $2\omega$ pulses with OAM by converting circularly polarized laser photons to second harmonic at the extremely sharp density gradients that exist at the sheath of a blown-out wake of a plasma accelerator. Recently, preliminary studies [32] have purported to show the spin-to-orbital angular momentum conversion during second harmonic generation in an underdense plasma. An a priori assumption in these works is that the total angular momentum (sum of SAM and OAM) is conserved during the $2\omega$ generation process. Although the conservation of OAM and SAM has been demonstrated in high harmonic generation in noble gases [33, 34], the conservation rule for the total angular momentum has yet to be systematically verified for plasmas.

As for the source of the second harmonic generation Ref. [32] assumes that the source must be due to density discontinuity at the plasma boundary. However, in an optically field ionized plasma, the ionization contour can be much wider than the laser spot size. Therefore, the intensity of the fundamental pump beam (which is close to the ionization threshold intensity) at the plasma edge can be far less than peak intensity. On the other hand, the intensity gradients can be very strong within the bulk of the plasma, providing another possible source for $2\omega$ generation [35]. In the experiments described here we use fundamental and higher-order Laguerre-Gaussian (LG) beams to produce plasmas by optical field ionization (OFI) of neutral helium gas. By observing the spatial distribution of the $2\omega$ photons from these plasmas we show that the strongest $2\omega$ emission corresponds well with the locations of the intensity gradients of the beams and not the density gradients of the plasma from ionization.

In this paper we show the conservation of the total optical angular momentum during the second harmonic generation in an underdense plasma. Whereas energy and linear momentum conservation demand that two fundamental photons are required to generate one $2\omega$ photon and that the $2\omega$ photons are emitted in the same direction as the fundamental photons, conservation of the total angular momentum requires the conversion of spin angular momentum of a circularly polarized (CP) light pulse into the orbital angular momentum of the second harmonic. By measuring the OAM of the $2\omega$ photons using different fundamental beam mode configuration we confirm this latter conjecture. Although plasma currents at $2\omega$ are the physical source of the emitted photons, the observed conservation rule implies that in the absence of any dissipation, plasma electrons do not retain any angular momentum and the plasma simply acts as a catalyst.

## Results

### SHG by pulses with angular momentum in underdense plasmas.

In the experiments reported here a short laser pulse with variable Laguerre-Gaussian spatial mode distribution and polarization (certain well known combinations of spin and orbital angular momentum) was focused to a "moderate" peak intensity of about $1.5 \times 10^{17}$ W cm$^{-2}$ inside a helium-filled chamber (Fig. 1). By moderate, we mean the intensity of the pulse is high enough to create a fully ionized region by a process known as optical field ionization in a finite region within and surrounding the laser spot but is well below the relativistic regime where the plasma electrons begin to oscillate close to the speed of light ($\sim 2 \times 10^{18}$ W cm$^{-2}$ for 800 nm). This way we minimize the relativistic effects on the one hand and a combination of relativistic self-focusing and ponderomotive force of the ultrashort laser pulse is not high enough to generate a significant transverse electron density depression on the other hand. The back-fill pressure of helium was set low (~$3.3 \times 10^{17}$ cm$^{-3}$) to avoid plasma-induced refraction of the pulse.

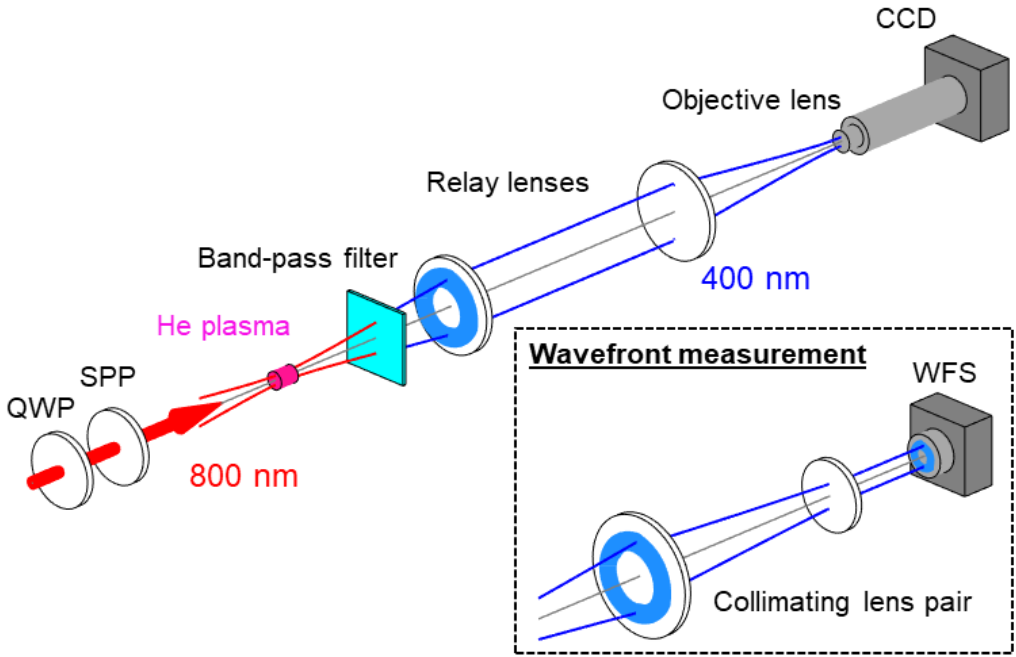

**Fig. 1 Experimental setup.** Schematic of set-up used to characterize the second-harmonic generation near-field and the far field (inset). Relay lens pair images the plasma exit to the objective lens. An ultrashort (50 fs (FWHM)), 800 nm laser pulse, containing up to 10 mJ of energy with controlled amount of SAM and OAM generated second harmonic light in plasma formed by OFI of static helium gas to give an electron density of $\sim 6.6 \times 10^{17}$ cm$^{-3}$ (double ionization of helium atoms). The $2\omega$ light (400 nm) was transmitted by a band-pass filter and transported out of the chamber. QWP: quarter wave plate; SPP: spiral phase plate. The imaging system was replaced by the diagnostics shown in the inset for wavefront measurement. In this setup, the $2\omega$ light was down collimated by a lens pair and then sent to a wavefront sensor (WFS).

As we mentioned earlier, second harmonic photons can be generated from a laser produced plasma when symmetry of the plasma medium is broken. In our experiments the plasma is underdense (laser frequency $\omega$ larger than the plasma frequency $\omega_p = (4\pi e^2 n_0/m)^{1/2}$, where $n_0$ is the ambient plasma density and $m$ is the electron mass). In such a plasma the dominant sources of nonlinearity that produce the second harmonic light are narrowed down to two first-order perturbation effects (Supplementary Note 1). Both effects come from the density perturbation $\delta n$ induced by the electron quiver motion in the laser field but they have different physical interpretations. The first is associated with the crossing of the quiver electrons in the ionization induced density gradients. In this case, the second harmonic current is directly proportional to the magnitude of the density gradient $\nabla n_0$. The second, is associated with quivering electrons in the intensity gradient where the perturbed electron trajectories lead to second harmonic emission. The first source is referred in this work as the density gradient (dg) contribution and the second as the intensity gradient (ig) contribution. Due to their distinctly different physical origins, two contributions can be discerned by the spatial distribution of the $2\omega$ photons within the plasma as we show in this section.

Consider the complex electric field of an optical Laguerre-Gaussian beam ($LG_p^l$) given by,

$$\mathbf{E} = (\hat{\mathbf{x}} + i\sigma\hat{\mathbf{y}})E_p^l(\rho)e^{i(kz-\omega t + l\phi)} \quad (1)$$

where $k$ is the wave number, $\rho = \sqrt{x^2 + y^2}/w_0$ is the normalized radial coordinate, $\phi = \tan^{-1}(y/x)$ is the azimuthal coordinate, $w_0$ is the focal spot size, and $E_p^l(\rho)$ is the transverse field envelope which is governed by the radial index $p$ and the azimuthal mode index $l$. The SAM value of the beam is determined by the polarization state, where $\sigma = 0$ for linear polarization (LP) and $\sigma = \pm 1$ for left-circular (LCP) and right-circular (RCP) polarization. The OAM value of the beam is determined by the azimuthal index $l = 0, \pm 1, \pm 2, \cdots$ (also known as the topological charge). For instance, when $p = 0$, we have an intensity profile described by $[E_0^l(\rho)]^2$, where $E_0^l(\rho) = E_0(\sqrt{2}\rho)^{|l|}\exp(-\rho^2)$ and $E_0$ is a constant.

If the local laser field is large enough to suppress the potential barrier of a bound electron in an atom/ion, this electron is released in a process generally referred to as optical field ionization. In the so-called barrier suppression ionization (BSI) limit this process will lead to an extremely sharp density gradient near the region where the laser exceeds the threshold intensity for ionizing a specific atom/ion [36]. For instance, in the present experiment we use He gas ionized by the laser. In this case, the density gradient induced current source that will produce $2\omega$ light is due to electrons quivering in two "shells" having large density gradients- the first between the region of the doubly and singly ionized He ions ($He^{2+}$-$He^{1+}$) and the second between the singly ionized He and neutral gas ($He^{1+}$-He). We denote the second harmonic current density from the density gradient contribution by $J_{2dg}$ and it is proportional to the magnitude of the density gradient and the local laser intensity. For a cylindrically symmetric intensity profile, we have

$$J_{2\text{dg}} \propto (\partial n_0/\partial \rho)[E_0^l(\rho_0)]^2 \tag{2}$$

where $\rho_0$ represents the position of the density jump.

Unlike the density gradient contribution that only depends on the local value of the intensity, the spatial distribution of the $2\omega$ current from the intensity gradient is modulated depending on the relative value of the beam's polarization (SAM) and helical phase front (OAM). This is because the quiver electron motion in the field gradient will be azimuthally perturbed when there is an additional azimuthal phase term $e^{il\phi}$. For circular polarization ($\sigma = \pm 1$), the amplitude of the second harmonic current density from the intensity gradient contribution (denoted by $J_{2\text{ig}}$) is given by

$$J_{2\text{ig}} \propto n_0[(|l| - \sigma l)\rho^{-1} - 2\rho][E_0^l(\rho)]^2 \tag{3}$$

where $\sigma = 1(-1)$ corresponds to LCP(RCP). When $l = 0$, $J_{2\text{ig}}$ has the same distribution as the intensity gradient.

Fortunately, these two mechanisms are separated in space. The spatial locations where the density and intensity gradient contributions are expected to be dominant respectively are shown in Fig. 2a for a fundamental, $l = 0$ mode LG beam. The current due to the intensity gradient contribution $J_{2\text{ig}}$ (blue curve) follows the intensity gradient (gray dashed curve) whereas the density gradient contribution is the largest near the positions where the BSI threshold intensities are reached for ionizing neutral He (brown dotted line) and He$^{1+}$ (green dotted line). Fig. 2b shows the measured 800 nm beam profile of this mode and Fig. 2c and 2d show the measured near field $2\omega$ images (color) and the expected positions of the of the largest intensity gradient (dashed circles) respectively for this case. There is an excellent agreement between the measured distributions of the $2\omega$ photons and the expected contour of the peak emission assuming the intensity gradient contribution is dominant. These $2\omega$ images have the expected annular shape from the intensity distribution but they have azimuthal inhomogeneities thought to arise from aberrations in the input beam. However, note that no detectable amount of $2\omega$ radiation is observed where the density gradients are the largest.

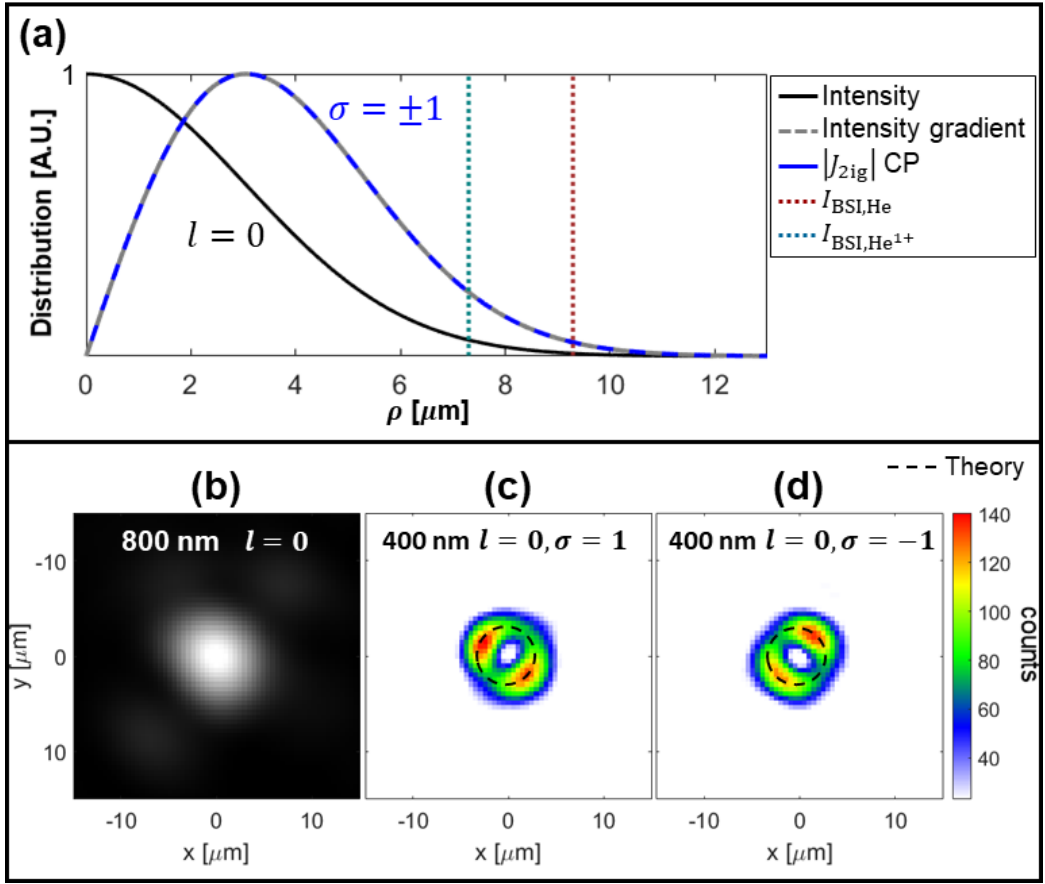

**Fig. 2 Spatial $2\omega$ current source and measured $2\omega$ photon distributions from $l = 0$, LG beam. a** Normalized radial distributions of the pump beam intensity (black), intensity gradient (dashed gray), and intensity-gradient-induced $2\omega$ current (blue) are calculated using CP laser pulses with a peak intensity of $1.5 \times 10^{17}$ W cm$^{-2}$. The radial coordinates where the laser intensity reaches the barrier suppression ionization threshold are marked for He (dotted brown) and He$^{1+}$ (dotted green) indicating the estimated position of the density gradients. **b** 800 nm focal spot image was used to estimate the focal spot size ($w_0 \sim 6$ μm) and calibrate the scale in calculations. $2\omega$ (400 nm) images from LCP and RCP are shown in **c**, **d** where the black dashed circles mark the estimated peak location deduced from the calculated radial distribution.

When $l \neq 0$, the radial distribution depends on the value $(|l| - \sigma l)$ (Eq. (3)) which is a quantity depending on both SAM $\sigma$ and OAM $l$. $J_{2ig}$ is calculated for $\sigma = \pm 1$ and $l = 1$ in Fig. 3a, where the distributions of the intensity (black curve) and intensity gradient (gray dashed curve) of the $LG_0^1$ are also shown. However, the distributions of $J_{2ig}$ for the LCP and the RCP cases are quite different now compared to those for the $l = 0$ shown in Fig 2. When $\sigma$ and $l$ have the same sign, a broad ring (red curve) is expected that has a radius larger than radius of the peak intensity position of the $LG_0^1$ mode. On the other hand, when $\sigma$ and $l$ have opposite signs, two rings are expected to appear – one on either side of the intensity peak. The stronger inner ring should be accompanied by a weaker outer ring in contrast to the previous case. This feature has been confirmed in the corresponding $2\omega$ images shown in Fig. 3c and 3d. The excellent agreement between the peak locations of the measured $2\omega$ emission and the calculated peak locations of the intensity gradients further confirms that the $2\omega$ photons are emitted from the plasma region where the laser intensity gradients are the largest rather than ionization induced density gradients.

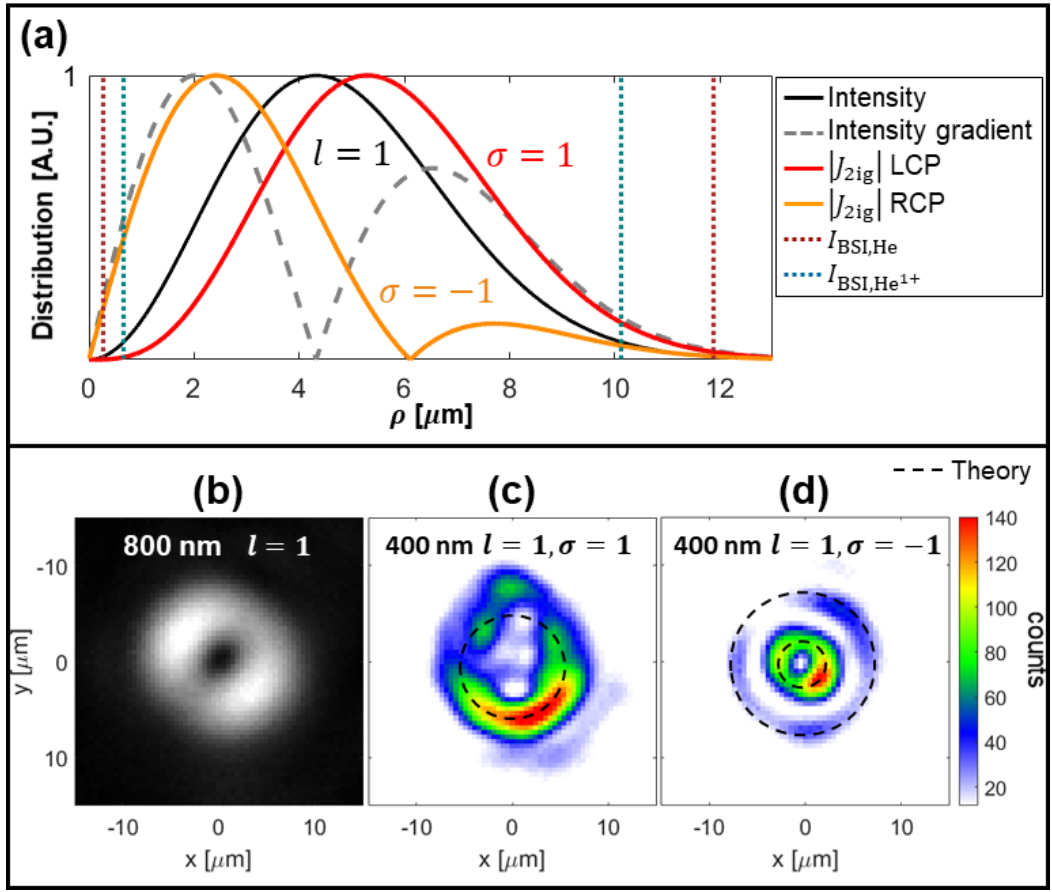

**Fig. 3 Spatial distribution of the $2\omega$ current from $LG_0^1$ beam. a** Normalized radial distributions of the pump beam intensity (black), intensity gradient (dashed gray), and intensity-gradient-induced $2\omega$ current (red and orange) are calculated using CP laser pulses with a peak intensity of $1.5 \times 10^{17}$ W cm$^{-2}$. BSI threshold intensities for He (dotted brown) and He$^{1+}$ (dotted green) are reached on both sides of the peak intensity. **b** Measured 800 nm $LG_0^1$ mode. Measured $2\omega$ (400 nm) image from LCP laser is shown in **c** where the dashed circle corresponds to the peak of the red curve in (**a**). Measured $2\omega$ image from RCP laser is shown in **d** where two dashed circles correspond to the inner and outer peaks of the orange curve in (**a**).

**Conservation rules of optical angular momentum.**

To determine the angular momentum properties of the output second harmonic beam, we calculate the far field electric field from the second harmonic current sources (see Methods section and Supplementary Note 2 for detail). The relevant terms of the $2\omega$ electric field generated from both mechanisms are identical, and therefore we have $\mathbf{E}_{2\omega,CP} \propto (i\cos\theta_m \hat{\boldsymbol{\theta}} \mp \hat{\boldsymbol{\phi}})\exp[i(2l \pm 2)\phi]$, where $\theta_m$ is the polar phase-match angle. When $\theta_m$ is small, the emission can be regarded as a plane wave propagating in the forward $z$ direction

$$\mathbf{E}_{2\omega,CP} \propto (\hat{\mathbf{x}} \pm i\hat{\mathbf{y}})\exp[i(2l \pm 1)\phi] \tag{4}$$

Eq. (4) shows the polarization state and the azimuthal phase terms of the $2\omega$ beam for a fundamental beam with SAM value $\sigma = \pm 1$ and OAM value $l$. By comparing the input and output beams, several conservation rules of optical angular momentum can be deduced. First, the polarization state is the same for small $\theta_m$. In other words, the spin angular momentum is preserved in the second-harmonic generation process,

$$s_{2\omega} = s_\omega = \sigma \tag{5}$$

where $s_\omega$ and $s_{2\omega}$ represent the SAM values of the fundamental and the $2\omega$ beams respectively. Second, the azimuthal phase factor of the $2\omega$ beam becomes $2l + \sigma$. This relation gives

$$l_{2\omega} = 2l_\omega + s_\omega \tag{6}$$

where $l_\omega$ and $l_{2\omega}$ are the azimuthal mode indexes of the fundamental and the $2\omega$ beam respectively. Using Eq. (5) and (6), rule for the conservation of the total angular momentum can be immediately obtained.

$$j_{2\omega} = 2j_\omega \tag{7}$$

where $j_n = l_n + s_n$ for respective harmonics $n$. Eq. (5)-(7) describe the conservation rules of spin, orbital, and total optical angular momentum in the second harmonic generation respectively.

Table 1 shows the combinations of angular momentum components of the fundamental and $2\omega$ beams based the conservation rules given by Eqns. (5), (6), and (7). For input beams with SAM values (-1, 0, 1) and OAM values (-1, 0, 1), these conservation rules give 5 combinations for the total, SAM and OAM of the $2\omega$ photons as shown in Table 1. Case A simply describes the situation where no optical angular momentum is involved. This is just a LP, Gaussian mode. Case B demonstrates simplest case that involves the conversion of spin-to-orbital angular momentum- a circularly polarized Gaussian pump beam. Case C shows the complete transfer of OAM from the fundamental to the $2\omega$ photons for a linearly polarized but $l = \pm 1$ pump beam. Cases D and E show the coupling of SAM and OAM in the second harmonic generation process. The SAM and OAM add up to either increase (case D) or decrease (case E) the orbital angular momentum.

| Case | Fundamental or Pump | | | SHG ($2\omega$) | | |
|---|---|---|---|---|---|---|
| | $j_\omega$ | $s_\omega$ | $l_\omega$ | $j_{2\omega}$ | $s_{2\omega}$ | $l_{2\omega}$ |
| A | 0 | 0 | 0 | 0 | 0 | **0** |
| B | $\pm 1$ | $\pm 1$ | 0 | $\pm 2$ | $\pm 1$ | **$\pm 1$** |
| C | $\pm 1$ | 0 | $\pm 1$ | $\pm 2$ | 0 | **$\pm 2$** |
| D | $\pm 2$ | $\pm 1$ | $\pm 1$ | $\pm 4$ | $\pm 1$ | **$\pm 3$** |
| E | 0 | $\mp 1$ | $\pm 1$ | 0 | $\mp 1$ | **$\pm 1$** |

**Table 1. Angular momentum conversion.** Combinations of angular momentum components of the fundamental and $2\omega$ beams, for the input fundamental or pump beams with linear or circular polarization ($\sigma=0, \pm 1$) and the azimuthal index $l=0, \pm 1$. The last column shows the expected $l$ of the $2\omega$ beam, which is the quantity measured in the experiments by measuring the spiral phase of the beam using two different methods.

**Measurement of the helical phase front of $2\omega$ light.**

    The experimental verification of the aforementioned conservation rules of optical angular momentum has been carried out by characterizing the helical phase of the second harmonic radiation. As shown in Table 1, by using various combinations of quarter-wave plate (QWP) and spiral phase plate (SPP) we can control the SAM and OAM respectively, thereby generating the fundamental or pump beam that has a known polarization and azimuthal mode index. We also show the expected values of the azimuthal index of the second harmonic light generated by the plasma for each setting of the QWP and the SPP. The QWP is set to make the output beam either LCP or RCP, corresponding to the SAM value of $+1$ or $-1$. The SPP is set to generate a Laguerre-Gaussian beam with $p = 0$ and $l = \pm 1$.

    The emerging second-harmonic radiation was measured using a commercial wavefront sensor (WFS-PHASICS SID4 HR) using the setup shown in the inset of Fig. 1. The sensor captures the far-field radiation intensity patterns (three examples are shown in Supplementary Fig. 2d-f) as well as the spatial phase information. Here we are more interested in the helical phase information of these intensity profiles in order to test the conservation laws for angular momentum. The helical phase structure is obtained from these recordings in the following way. For example, the $2\omega$ phase recorded by the WFS using LCP and RCP LG laser pulses are subtracted to give the phase information (see Methods section for more information). This procedure leads to an azimuthally increasing relative phase distribution as shown in Fig. 4. The magnitude of the total phase change is

$2\pi(l_1 - l_2)$ radians if two beams have azimuthal indexes of $l_1$ and $l_2$. Since in this case we compare two cases of beams with opposite sense of rotation (helicity), $l_2 = -l_1$, we expect twice the phase shift, which leads to a total phase shift of $4\pi|l_{2\omega}|$ radians.

Fig 4a shows the azimuthal phase obtained when a circularly polarized fundamental LG beam was used. This is Case B in Table 1. Although the pump beam has no OAM we expect the $2\omega$ beam to have $l_{2\omega} = \pm 1$ and a total azimuthal phase of $4\pi$ radians as seen in Fig. 4a. Fig. 4b shows the phase shift distribution when the $2\omega$ photons from LP LG beams with $l_\omega = \pm 1$ are measured. The expected magnitude of total relative phase shift (Case C in Table 1) is now $8\pi$ radians if there were twice the amount of OAM in the $2\omega$ beam ($l_{2\omega} = \pm 2$) indicating OAM was completely transferred from the fundamental to the second harmonic. Now we discuss what happens when the SPP was used in tandem with the QWP (for both case D and E in Table 1). In Fig. 4c, OAM and SAM values of the input beams were set according to case D. A total relative phase shift of $12\pi$ radians is expected and observed since the $2\omega$ output beam possesses $l_{2\omega} = \pm 3$. Even though this method only demonstrated a relative helical phase shift, the total phase shift implies the amount OAM carried by the output beams are as predicted in Table 1.

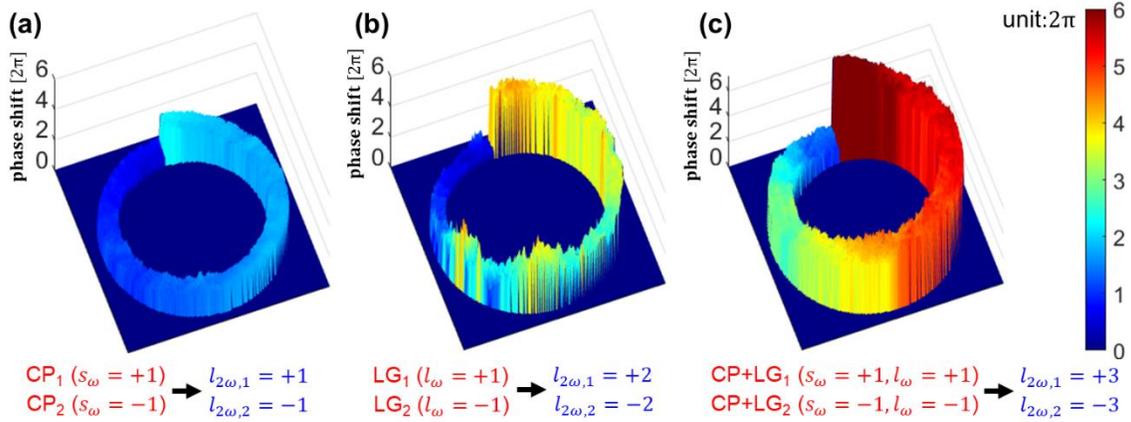

**Fig. 4 Spiral phase measured by the wave-front sensor.** Relative helical phase distributions between two shots from **a** LCP and RCP beams ($CP_1$-$CP_2$), **b** linearly polarized LG beams with opposite helicities ($LG_1$-$LG_2$), **c** circularly polarized LG beams with opposite helicities (($CP+LG)_1$-$(CP+LG)_2$).

A second technique was used to confirm the results of the helical phase measurement obtained using the WFS. In this set-up the collimated second harmonic light was sent through a double-slit interferometer for a direct measurement of the azimuthal mode index of the beams (Fig. 5a). The beam passed through the center of a custom-made double-slit target and generated an interference pattern. A double slit mask with a slit width of 30 $\mu$m and a separation of 2.8 mm was used to retrieve the difference in the spatial phase from a range of azimuthal coordinates intercepted by the slits. A screen was placed about 1 m behind the mask. Contrary to the typical straight fringe pattern obtained when a plane wave traverses the two slits in Young's double slit experiment, a beam with a helical wavefront yields a kinked fringe pattern bent in a direction according to its helicity [37]. This effect is clearly demonstrated in Fig. 5. For a linear polarized Gaussian beam (case A), its radiation pattern has two lobes along the polarization direction (Fig. 5b) and its interference pattern yields straight line fringes (Fig. 5c) showing that the output $2\omega$ beam carries no OAM. For a circularly polarized Gaussian beam (case B, $s_\omega = \pm 1$), annular emission pattern (Fig. 5d) appears. Fig. 5e and 5f are results of double slit experiments from the LCP and RCP cases respectively. The fringe patterns for two cases now have kinks (bends) in opposite direction with nearly equal amount of shift.

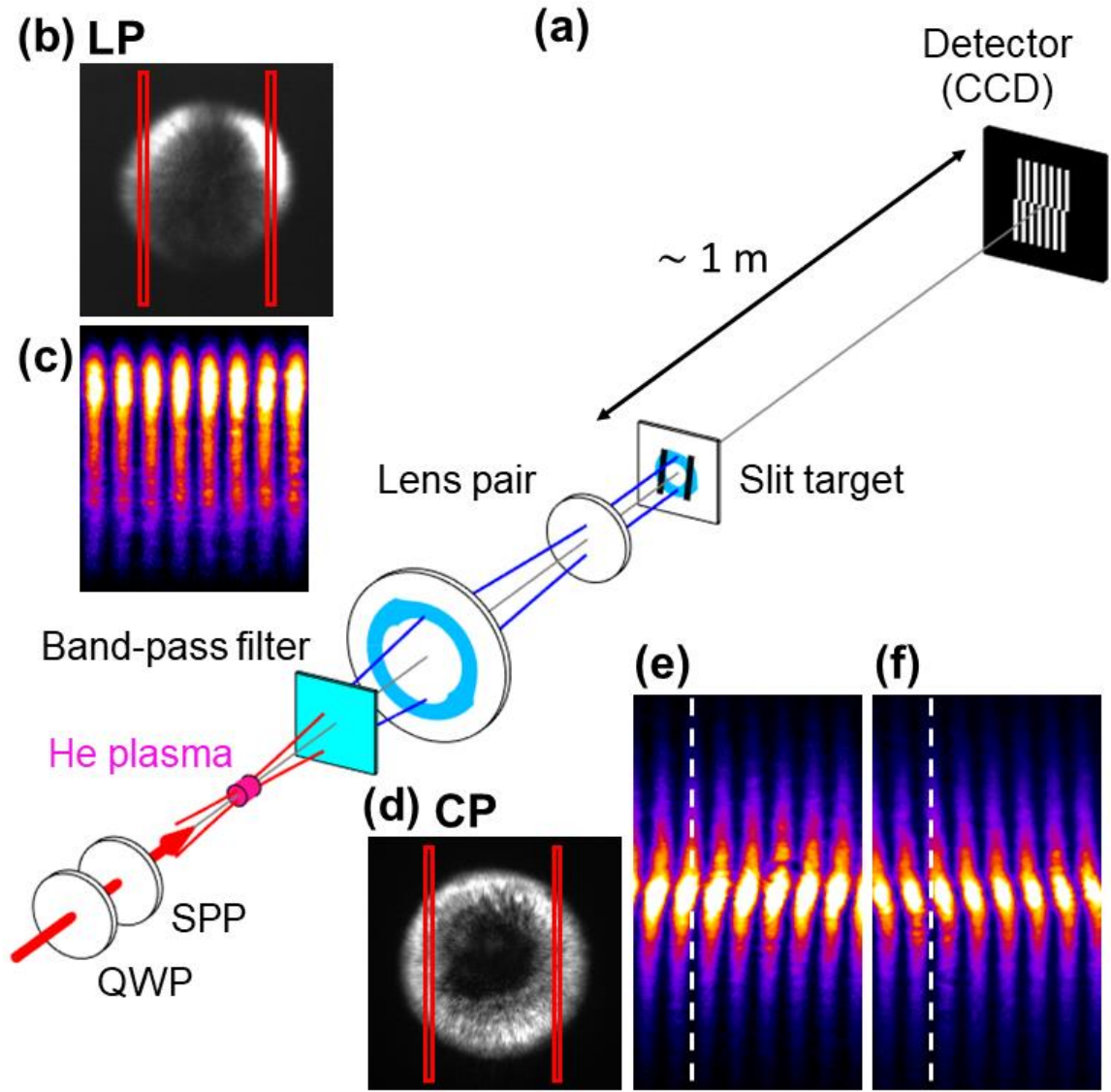

**Fig. 5 Double-slit interferometry. a** Schematic setup of the interferometer. **b** 2ω radiation pattern from a linearly polarized beam and **c** the corresponding fringe pattern. **d** 2ω radiation pattern from a circularly polarized beam. **e** and **f** show the fringe patterns for LCP and RCP cases. The red rectangles in **b** and **d** indicate the approximate position of the slits. The white dashed line in **e** and **f** shows the relative bending of the fringes.

Single-shot measurements of the interference pattern using different polarizations and spatial modes of the $2\omega$ beams are shown in Fig. 6a for the cases B, C, D, and E listed in Table 1. For different cases, different magnitudes of shift of the fringes were observed. Fig. 6b shows a series of calculated fringe patterns (synthetic interferograms) by using the slit width and the slit-screen distance in the experiment and assuming a homogeneous input light with different index $l$. The exact value of OAM carried by the beam can be evaluated measuring the shift for a fringe from the top to bottom and by comparing the experimental interference pattern with the synthetic pattern. Contrary to the calculated results, the actual patterns have inhomogeneous intensity distributions due to the local variations of intensity on the slits. For cases shown in Fig. 6, the estimated $l_{2\omega}$ are $-1.00 \pm 0.04$ ($-1$), $1.85 \pm 0.07$ ($+2$), $3.11 \pm 0.04$ ($+3$), and $1.03 \pm 0.04$ ($+1$) where the predicted values are in the parentheses. The agreement is good. (See Supplementary Note 3 for the discussion of the error bars.) Similar magnitudes of shift between the experimental and calculated fringes are also demonstrated by the white and red dashed lines in Fig. 6a and 6b. These results are strong indications that the conservation law for OAM given by Eq. (6) is obeyed.

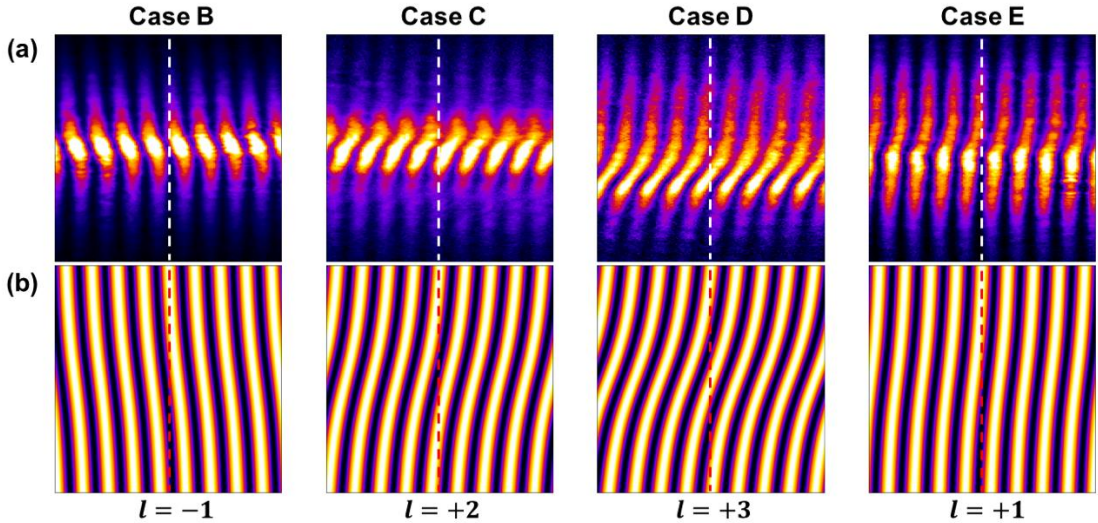

**Fig 6. Interference patterns from spiral phase fronts. a** Observed and **b** calculated interference fringe patterns for different cases in Table 1. The input beam parameters from left to right: Case B ($s = -1, l = 0$), Case C ($s = 0, l = +1$), Case D ($s = +1, l = +1$), Case E ($s = -1, l = +1$). The white and red dash lines in **a** and **b** overlap with one of the bright fringes at the top for each case B-E for measuring the phase shift. The experimental patterns shown were selected from sets of 10 consecutive shots (movies can be found in the supplemental material).

## Discussion

The results from two different diagnostics indicate that the helical phase structure and the OAM mode index of the second harmonic radiation emitted from an optical-field ionized helium plasma matches the value predicted by the conservation rule Eq. (6) for the input SAM and OAM values of 0 or $\pm 1$. It is then reasonable to suppose that the conservation rules can be extended to other integer OAM values $l = \pm 2, \pm 3, \pm 4, \cdots$. Therefore, variable orbital angular momentum can be generated this way using the spin-to-orbital angular momentum conversion. However, small but definite changes of polarization between the fundamental and $2\omega$ beams were observed (Supplementary Note 4). This can give rise to a fractional value of spin angular momentum. There are physical mechanisms that may depolarize the $2\omega$ radiation in the plasma itself. First, it is well known that the light with angular momentum can generate axial magnetic field in plasmas through inverse Faraday effect [38, 23]. Axial magnetic field can depolarize the radiated $2\omega$ emission via the Faraday effect. Second, laser-ionized plasmas are known to have non-thermal electron velocity distributions and are susceptible to plasma kinetic instabilities [39, 40]. The axial helicoidal magnetic field induced by electron filamentation/Weibel instability can cause depolarization of the $2\omega$ radiation and plasma emission. Third, refraction from the density gradients ($He^{1+}$-$He^{2+}$ and $He^{1+}$-$He$) can rotate the polarization plane for a fraction of emitted $2\omega$ light and cause depolarization. It should be noted that the coherent nature of the single-shot double slit measurement provides a better measurement than the azimuthal phase deduced from the WFS which requires subtraction of two independently obtained shots that have equal but opposite mode index. The agreement between the experimental results and theoretical predictions provides strong support that Eq. (7) is valid for our experimental parameters.

In conclusion, the source and the conservation of total optical angular momentum in the second harmonic generation from optical-field ionized plasmas has been verified experimentally. The conservation law is accompanied by conversion of the spin-to-orbital angular momentum. The results agree with the conservation rules of angular momentum, which have been derived using the simple electron quiver motion model. In the model, free electrons absorb both spin and orbital angular momentum from the input beam and move accordingly. When these electrons pass through density or intensity gradients, second harmonic currents with the same polarization and phase factor of the beam are generated and angular momentum is transferred to the resultant radiation. In our experiment $2\omega$ radiation originates from plasma regions in the vicinity of the largest intensity gradients of the pump laser. We note that the present mechanism for the current source assumes no dissipation (collisions) or coupling to a plasma mode and the plasma electrons simply catalyze this conversion of the OAM in the harmonic generation process unlike in the case of parametric instabilities where energy, momentum, and angular momentum conservation involves a collective mode of the plasma.

## Methods

**Calculation of far-field radiation.** Following a two-step procedure described in reference [41], the far-field radiation field can be calculated systematically from a known current source. In the first step, the vector potential is found from the integration

$$\mathbf{A}_{2\omega} = \iiint_V dv' \frac{\mathbf{J}_{2\omega}(\rho',\phi',z')}{c} \frac{e^{ik|\mathbf{r}-\mathbf{r}'|}}{|\mathbf{r}-\mathbf{r}'|}$$

in far-field approximation. In the second step, the electric field $\mathbf{E}_{2\omega}$ is calculated from the vector potential $\mathbf{A}_{2\omega}$. The integration over the current density is done in cylindrical coordinates, but the components of the vector potential and the field are solved in spherical coordinates.

For the density gradient contribution, because of the steep density change induced by optical-field ionization, it is simpler to assume a region of density gradient as a delta function at $\rho = \rho_0$. Substitute $\partial n_0/\partial \rho = \pm C_g \delta(\rho - \rho_0)$ into Eq. (2), where $C_g$ is a constant. The electric field distribution from the density gradient is found to be

$$\mathbf{E}_{2\omega dg} \propto \left(i\cos\theta\,\hat{\boldsymbol{\theta}} \mp \hat{\boldsymbol{\phi}}\right) \text{sinc}[(k_2\cos\theta - 2k_1)z_R] \qquad (8)$$
$$\times B_{2,l\pm 1}(k_2\rho_0 \sin\theta) e^{i[k_2 r - 2\omega t + 2(l\pm 1)\phi]}$$

where $B_{2l\pm 1}$ is the Bessel function of the first kind of order $2l \pm 1$ and $z_R$ is the Rayleigh range. Note that each density jump in plasma can be calculated separately using Eq. (8).

For the intensity gradient contribution, integration for Eq. (3) can be done analytically (Supplementary Note 2). The electric field distribution from the intensity gradient can be written as

$$\mathbf{E}_{2\omega ig} \propto \left(i\cos\theta\,\hat{\boldsymbol{\theta}} \mp \hat{\boldsymbol{\phi}}\right) \text{sinc}[(k_2\cos\theta - 2k_1)z_R] e^{i[k_2 r - 2\omega t + 2(l\pm 1)\phi]} \Psi_{l,\pm 1}(\theta) \qquad (9)$$

where $\Psi_{l,\pm 1}(\theta)$ is a function of $\theta$ and $\pm 1$ correspond to LCP/RCP. $\Psi_{l,\pm 1}(\theta)$ has the similar feature of high-order Bessel functions and also yields a pattern with multiple rings.

**Experimental setup.** The experimental setup is shown schematically in Fig. 1 and 5a. The pump laser pulse has a central wavelength of 800 nm and a pulse duration of ~50 fs (full width half maximum). The vacuum chamber was filled with a static fill-pressure of 10 torr of He gas. Helium was ionized by focusing the pump beam by an off-axis parabolic mirror (OAP) with a f-number of 12. To ensure a full ionization at the focus, different pump energies were used for different input modes to keep the peak intensity larger than $1.5 \times 10^{17}$ W cm$^{-2}$, which is far beyond the ionization threshold of He$^{1+}$ ion. A blue glass (Newport BG40) was placed after the focus to block most of the 800 nm pump beam energy but allow the transmission of the 400 nm beam. The divergent second harmonic radiation was then collimated by a lens pair and was guided into different detection systems. For most measurements, another band-pass filter (centered at 406 nm with a bandwidth of 40 nm) was inserted in front of the detector to block stray light and plasma line emission. SAM and OAM of the input beam were introduced by a zero-order quarter-wave plate and a spiral phase plate. Since the incident beam had a top-hat (super-Gaussian) intensity profile, the output beam of the SPP is better described as a modified Laguerre-Gaussian beam. The modified LG beam has slightly different intensity profile at the focal plane compared with the ideal mode [31].

**Detail of the spiral phase measurement.** Shack-Hartmann wavefront sensors have been shown to be useful in reconstruction of the helical phase of high-harmonic generation with optical vortices [42, 43] despite difficulties related to phase dislocation inherent in optical vortices. Numerical methods are often required to avoid the discontinuities in the phase front [42, 44]. In this work, we applied annular numerical masks matching the $2\omega$ beam profiles (Supplementary Fig. 2d-f) and took the local wavefront information with a plane wave reference for each shot. The recorded interferograms (phase distribution between the signal $2\omega$ beam and a plane wave reference) with opposite signs of helicity were subtracted to reconstruct a relative phase distribution. The main reason for using a relative measurement is to exclude the defocus and spherical aberration of the $2\omega$ beam, which is constantly larger than $2\pi$ rad. The spiral phase information is present together with additional phase information that is sensitive to azimuthal index and laser conditions. Since two beams with the same $|l|$ value have similar spiral phase profiles, the subtraction of two images can mitigate the additional phase information. What remains will be the differential spatial phase that reveals the spiral phase structure of vortex beams. The magnitude of the total phase increment is $2\pi(l_1 - l_2)$ radians if two beams have azimuthal index of $l_1$ and $l_2$.

The WFS method has some issues when the spiral phase is being extracted from the beam directly, so it has to be inferred from a relative phase measurement. The double slit interferometer has no such problem. The azimuthal phase can be obtained in single shot but this method unlike the WFS method is not able to give the phase distribution of the whole beam. In this respect the two methods complement one another. The calculated double-slit interference patterns are generated using the intensity profile given by $I(x,y) = I_0 \cos^2(\pi x a / \lambda L + \Delta\phi(y)/2)$ where $I_0$ is the normalized peak intensity, $a$ is the distance between two slits, $\lambda$ is the wavelength of the input light, $L$ is the distance between the slits and the screen, and $\Delta\phi(y)$ is the phase difference between two slits along the vertical direction y. The vertical phase shift term $\Delta\phi(y)$ is proportional to azimuthal index of the input light but modulated by $a$. It could be difficult to precisely determine arbitrary azimuthal index due to the inhomogeneous light intensity distribution [45]. However, one can estimate this value using $I(x,y)$ and mitigate the effect of inhomogeneity by using average amount of shift in the estimation.

Both methods were calibrated using 800 nm beams that carried known azimuthal indexes ($l = 0, \pm 1$) and gave the expected results. Measurement errors at $2\omega$ are mainly due to shot to shot variations of the wavefront aberrations.

## Acknowledgements

This work was supported by DOE grant DE-SC0010064, NSF grant 1734315, AFOSR grant FA9550-16-1-0139, and ONR MURI award N00014-17-1-2705. We thank Professor W. B. Mori for useful discussions.

## Reference

[1] Allen, L., Beijersbergen, M. W., Spreeuw, R. J. C. & Woerdman, J. P. Orbital angular momentum of light and the transformation of Laguerre-Gaussian laser modes. *Phys. Rev. A* **45,** 8185-8189 (1992).

[2] Beth, R. A. Mechanical detection and measurement of the angular momentum of light. *Phys. Rev.* **50,** 115-125 (1936).

[3] He, H., Friese, M. E. J., Heckenberg, N. R. & Rubinsztein-Dunlop, H. Direct observation of transfer of angular momentum to absorptive particles from a laser beam with a phase singularity. *Phys. Rev. Lett.* **75,** 826-829 (1995).

[4] O'Neil, A. T., MacVicar, I., Allen, L. & Padgett, M. J. Intrinsic and extrinsic nature of the orbital angular momentum of a light beam. *Phys. Rev. Lett.* **88,** 053601 (2002).

[5] Friese, M. E. J., Enger, J., Rubinsztein-Dunlop, H. & Heckenberg, N. R. Optical angular-momentum transfer to trapped absorbing particles. *Phys. Rev. A* **54,** 1593-1596 (1996).

[6] Simpson, N. B., Dholakia, K., Allen, L. & Padgett, M. J. Mechanical equivalence of spin and orbital angular momentum of light: an optical spanner. *Opt. Lett.* **22,** 52-54 (1997).

[7] Courtial, J., Robertson, D. A., Dholakia, K., Allen, L. & Padgett, M. J. Rotational frequency shift of a light beam. *Phys. Rev. Lett.* **81,** 4828-4830 (1998).

[8] Padgett, M. J. & Bowman, R. Tweezers with a twist. *Nat. Photonics* **5,** 343-348 (2011).

[9] Wang, W. P. et al. New Optical manipulation of relativistic vortex cutter. *Phys. Rev. Lett.* **122,** 024801 (2019).

[10] Bomzon, Z., Biener, G., Kleiner, V. & Hasman, E. Space-variant Pancharatnam–Berry phase optical elements with computer-generated subwavelength gratings. *Opt. Lett.* **27,** 1141-1143 (2002).

[11] Marrucci, L., Manzo, C. & Paparo, D. Optical Spin-to-Orbital Angular Momentum Conversion in Inhomogeneous Anisotropic Media. *Phys. Rev. Lett.* **96,** 163905 (2006).

[12] Karimi, E., Schulz, S. A., De Leon, I., Qassim, H., Upham, J. & Boyd, R. W. Generating optical orbital angular momentum at visible wavelengths using a plasmonic metasurface. *Light Sci Appl* **3**, e167 (2014).

[13] Devlin, R. C. et al. Spin-to-orbital angular momentum conversion in dielectric metasurfaces. *Opt. Lett.* **25,** 377-393 (2017).

[14] Kruer, W. L. *The Physics of Laser Plasma Interactions* (CRC Press, Boca Raton, 2003).

[15] Eliezer, S. *The Interaction of High-power Lasers with Plasmas* (CRC Press, Boca Raton, 2002).

[16] Mendonça, J. T., Thide, B. & Then, H. Stimulated Raman and Brillouin backscattering of collimated beams carrying orbital angular momentum. *Phys. Rev. Lett.* **102,** 185005 (2009).

[17] Vieira, J., Mendonça, J. T. & Quéré, F. Optical Control of the Topology of Laser-Plasma Accelerators. *Phys. Rev. Lett.* **121,** 054801 (2018).

[18] Vieira, J. & Mendonça, J. T. Nonlinear laser driven donut wakefields for positron and electron acceleration. *Phys. Rev. Lett.* **112,** 215001 (2014).

[19] Mendonça, J. T. & Vieira, J. Donut wakefields generated by intense laser pulses with orbital angular momentum. *Phys. Plasmas* **21,** 033107 (2014).

[20] Zhang, G. B. et al. Acceleration and evolution of a hollow electron beam in wakefields driven by a Laguerre-Gaussian laser pulse. *Phys. Plasmas* **23**, 033114 (2016).

[21] Brabetz, C. et al. Laser-driven ion acceleration with hollow laser beams. *Phys. Plasmas* **22,** 013105 (2015).

[22] Luís Martins, J., Vieira, J., Ferri, J. & Fülöp, T. Radiation emission in laser-wakefelds driven by structured laser pulses with orbital angular momentum. *Sci. Rep.* **9,** 9840; 10.1038/s41598-019-45474-8 (2019).

[23] Ali, S., Davies, J. R. & Mendonça, J. T. Inverse Faraday effect with linearly polarized laser pulses. *Phys. Rev. Lett.* **105,** 035001 (2010).

[24] Shi, Y. et al. Magnetic Field generation in plasma waves driven by copropagating intense twisted lasers generation of intense high-order vortex harmonics. *Phys. Rev. Lett.* **121,** 145002 (2018).

[25] Gordon, D. F., Hafizi, B. & Ting A. Nonlinear conversion of photon spin to photon orbital angular momentum. *Opt. Lett.* **34,** 3280-3282 (2009).

[26] Qu, K., Jia, Q. & Fisch, N. J. Plasma q-plate for generation and manipulation of intense optical vortices. *Phys. Rev. E* **96,** 053207 (2017).

[27] Zhang, X. et al. Generation of intense high-order vortex harmonics. *Phys. Rev. Lett.* **114,** 173901 (2015).

[28] Leblanc, A. et al. Plasma holograms for ultrahigh-intensity optics. *Nat. Phys.* **13,** 440-443 (2017).

[29] Vieira, J. et al. Amplification and generation of ultra-intense twisted laser pulses via stimulated Raman scattering. *Nat. Commun.* **7,** 10371; 10.1038/ncomms10371 (2016).

[30] Denoeud, A., Chopineau, L., Leblanc, A. & Quere, F. Interaction of ultraintense laser vortices with plasma mirrors. *Phys. Rev. Lett.* **118,** 033902 (2017).

[31] Ohland, J. B., Eisenbarth, U., Roth, M. & Bagnoud, V. A study on the effects and visibility of low-order aberrations on laser beams with orbital angular momentum. *Appl. Phys. B* **125,** 202 (2019).

[32] Wilhelm, A., Schmidt, D. & Durfee, C. Nonlinear optical conversion of photon Spin to orbital angular momentum. in *Conference on Lasers and Electro-Optics*, OSA Terchnical Digest (Optical Society of America, 2018), paper FTh3E. 3.

[33] Gariepy, G. et al. Creating high-harmonic beams with controlled orbital angular momentum. *Phys. Rev. Lett.* **113,** 153901 (2014).

[34] Dorney, K. M. et al. Controlling the polarization and vortex charge of attosecond high-harmonic beams via simultaneous spin–orbit momentum conservation. *Nat. Photonics* **13,** 123-130 (2019).

[35] Mori, M., Takahashi, E. & Kondo, K. Image of second harmonic emission generated from ponderomotively excited plasma density gradient. *Phys. Plasmas* **9,** 2812-2815 (2002).

[36] Augst, S., Meyerhofer, D. D., Strickland, D. & Chin, S. L. Laser ionization of noble gases by Coulomb-barrier suppression. *J. Opt. Soc. Am. B* **8,** 858-867 (1991).

[37] Sztul, H. I. & Alfano, R. R. Double-slit interference with Laguerre–Gaussian beams. *Opt. Lett.* **31,** 999-1001 (2006).

[38] Deschamps, J., Fitaire, M. & Lagoutte, M. Inverse Faraday effect in a plasma. *Phys. Rev. Lett.* **25,** 1330-1332 (1970).

[39] Zhang, C. -J., Huang, C.-K., Marsh, K. A., Clayton, C. E., Mori, W. B. & Joshi, C. Ultrafast optical field–ionized gases—A laboratory platform for studying kinetic plasma instabilities. *Sci. Adv* **5**(9):eaax4545 (2019).

[40] Huang, C.-K., Zhang, C. -J., Marsh, K. A., Clayton, C. E. & Joshi, C. Initializing anisotropic electron velocity distribution functions in optical-field ionized plasmas. *Plasma Phys. Control. Fusion* **62,** 024001 (2020).

[41] Balanis, C. A. *Advanced Engineering Electromagnetics* (Wiley, New York, 2012).

[42] Gauthier, D. et al. Tunable orbital angular momentum in high-harmonic generation. *Nat. Commun.* **8,** 14971; 10.1038/ncomms14971 (2017).

[43] Ribič, P. R. et al. Extreme-ultraviolet vortices from a free-electron laser. *Phys. Rev. X* **7,** 031036 (2017).

[44] Starikov, F. A. et al. Wavefront reconstruction of an optical vortex by a Hartmann–Shack sensor. *Opt. Lett.* **32,** 2291-2293 (2007).

[45] Emile, O. & Emile, J. Young's double-slit interference pattern from a twisted beam. *Appl. Phys. B* **117,** 487-491 (2014).